\documentclass[8.5pt,twoside,twocolumn]{article}
\usepackage[super,sort&compress,comma]{natbib} 
\usepackage{mhchem}
\usepackage{times}
\usepackage{sectsty}
\usepackage{balance} 

\usepackage{amsmath}
\usepackage{amssymb}
\usepackage{amsthm}

\usepackage{graphicx} 
\usepackage{lastpage}
\usepackage[format=plain,justification=raggedright,singlelinecheck=false,font=small,labelfont=bf,labelsep=space]{caption} 
\usepackage{fancyhdr}
\begin{document}

\thispagestyle{plain}
\fancypagestyle{plain}{
\fancyhead[L]{\includegraphics[height=8pt]{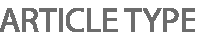}}
\fancyhead[C]{\hspace{-1cm}\includegraphics[height=20pt]{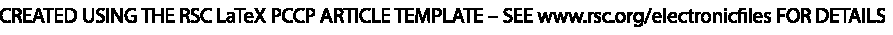}}
\fancyhead[R]{\includegraphics[height=10pt]{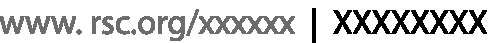}\vspace{-0.2cm}}
\renewcommand{\headrulewidth}{1pt}}
\renewcommand{\thefootnote}{\fnsymbol{footnote}}
\renewcommand\footnoterule{\vspace*{1pt}%
\hrule width 3.4in height 0.4pt \vspace*{5pt}} 
\setcounter{secnumdepth}{5}

\makeatletter 
\def\subsubsection{\@startsection{subsubsection}{3}{10pt}{-1.25ex plus -1ex minus -.1ex}{0ex plus 0ex}{\normalsize\bf}} 
\def\paragraph{\@startsection{paragraph}{4}{10pt}{-1.25ex plus -1ex minus -.1ex}{0ex plus 0ex}{\normalsize\textit}} 
\renewcommand\@biblabel[1]{#1}            
\renewcommand\@makefntext[1]%
{\noindent\makebox[0pt][r]{\@thefnmark\,}#1}
\makeatother 
\renewcommand{\figurename}{\small{Fig.}~}
\sectionfont{\large}
\subsectionfont{\normalsize} 

\fancyfoot{}
\fancyfoot[LO,RE]{\vspace{-7pt}\includegraphics[height=9pt]{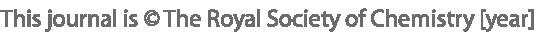}}
\fancyfoot[CO]{\vspace{-7.2pt}\hspace{12.2cm}\includegraphics{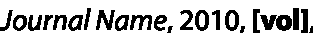}}
\fancyfoot[CE]{\vspace{-7.5pt}\hspace{-13.5cm}\includegraphics{RF.eps}}
\fancyfoot[RO]{\footnotesize{\sffamily{1--\pageref{LastPage} ~\textbar  \hspace{2pt}\thepage}}}
\fancyfoot[LE]{\footnotesize{\sffamily{\thepage~\textbar\hspace{3.45cm} 1--\pageref{LastPage}}}}
\fancyhead{}
\renewcommand{\headrulewidth}{1pt} 
\renewcommand{\footrulewidth}{1pt}
\setlength{\arrayrulewidth}{1pt}
\setlength{\columnsep}{6.5mm}
\setlength\bibsep{1pt}

\newcommand{\fat}[1]{\mathbf{#1}}
\newcommand{\fatsym}[1]{\boldsymbol{#1}}
\newcommand{\D}{\mbox{d}}

\newcommand{\F}{\mathcal{F}}
\newcommand{\var}{\delta}

\twocolumn[
  \begin{@twocolumnfalse}
\noindent\LARGE{\textbf{Pattern formation of dipolar colloids in rotating fields: Layering and synchronization.$^\dag$}}
\vspace{0.6cm}

\noindent\large{\textbf{Sebastian J\"ager\textit{$^{a}$} and Sabine H. L. Klapp$^{\ast}$}\textit{$^{a}$}}\vspace{0.5cm}

\noindent\textit{\small{\textbf{Received Xth XXXXXXXXXX 20XX, Accepted Xth XXXXXXXXX 20XX\newline
First published on the web Xth XXXXXXXXXX 200X}}}

\noindent \textbf{\small{DOI: 10.1039/b000000x}}
\vspace{0.6cm}

\noindent \normalsize{We report Brownian dynamics (BD) simulation and
theoretical results for a system of spherical colloidal particles with
permanent dipole moments in a rotating magnetic field. Performing simulations
at a fixed packing fraction and dipole coupling parameter, we construct a full
non-equilibrium phase diagram as function of the driving frequency ($\omega_0$)
and field strength ($B_0$). This diagram contains both synchronized states,
where the individual particles follow the field with (on average) constant
phase difference, and asynchronous states. The synchronization is accompanied
by layer formation, \textit{i.e.} by spatial symmetry-breaking, similar to
systems of induced dipoles in rotating fields. In the permanent-dipole case,
however, too large $\omega_0$ yield a breakdown of layering, supplemented by
complex changes of the single-particle rotational dynamics from synchronous to
asynchronous behavior. We show that the limit frequencies $\omega_c$ can be
well described as a bifurcation in the nonlinear equation of motion of a single
particle rotating in a viscous medium. Finally, we present a simple density
functional theory, which describes the emergence of layers in perfectly
synchronized states as an equilibrium phase transition. } \vspace{0.5cm}
\end{@twocolumnfalse} ]

\section{Introduction}
\footnotetext{\textit{$^{a}$~Institute of Theoretical Physics, Technical
University Berlin, Hardenbergstr. 36, 10623 Berlin, Germany. E-mail:
klapp@physik.tu-berlin.de}}

The dynamics of anisotropic particles driven by time-dependent, magnetic or
electric, external fields is currently a topic receiving much attention. Many
experimental and theoretical studies in this area focus on the field-induced
dynamics of a {\em isolated} nanoparticle such as a magnetic
rod,\cite{Tierno2009,Coq2010,Dhar2007} a magnetic chain \cite{casic} or
filament,\cite{Dreyfus2005} or an optically excitable nanorod \cite{shelton} in
a viscous medium. Understanding the resulting single-particle rotational
dynamics is particularly important for actuators,\cite{Coq2010} molecular
switches, particles in optical traps,\cite{shelton} and in the more general
context of microfluidics.\cite{Dhar2007} From the theoretical side, these
problems are often successfully analyzed on the basis of single-particle,
nonlinear equations for the driven rotational motion in the presence of
solvent-induced friction.\cite{Tierno2009,Coq2010,Dhar2007,shelton} Typically,
the particle dynamics exhibits a ``linear'' regime at low driving frequencies,
where the particle axis follows the field, and various types of nonlinear
behavior at high frequencies, such as rotation against the
torque.\cite{shelton} Many of these nonlinear phenomena, including transient
behavior such as conformal transitions\cite{casic} of magnetic chains following
a sudden switch-on of the driving field, can also be observed
experimentally.\cite{Tierno2009,Dhar2007}

Apart from the single-particle dynamics, another current focus concerns the
{\em self-assembly} behavior in colloidal many-particle systems that are
exposed to rotating fields. Indeed, in material science, time-dependent fields
are currently realized as a powerful tool to control self-assembly processes,
which are an important prerequisite for synthesizing functional
materials.\cite{leunissen, Douglas2010} A classical example in this context,
first discussed by Martin {\em et al.},\cite{martin2} are systems of
paramagnetic (or polarizable) spherical particles in magnetic (electric) fields
rotating in a plane. For sufficient field strength, both experiment and
computer simulations \cite{martin2,martin6,elsner} reveal the formation of
layers in the field plane, \textit{i.e.} a {\em spatial} symmetry breaking
induced by the rotating field. Indeed, a rotating in-plane field generates, on
averaging over time, an inverted dipolar pair interaction with in-plane
attraction and repulsion along the rotation axis.\cite{martin1,elsner}
Therefore, the structures induced by planar rotating fields are markedly
different from those observed in a consant and homogeneous field, which
supports the formation of field-aligned chains (low
densities)\cite{philip,butter,Jordanovic2011} and bulk
crystals.\cite{dassanayake,hynninen}

The general idea to use time-dependent fields to tune pair interactions and
thereby control the morphology of self-assembled structures has meanwhile
become more and more popular (see Refs.~\citenum{martin5,martin_strange}), a
recent example being the formation of self-healing membranes of
superparamagnetic particles in tilted rotating fields. \cite{ostermann}
Interestingly, these self-assembly phenomena can often be explained from an
equilibrium perspective involving the free energy and resulting phase behavior
of a many-particle system in a time-averaged field. Clearly, the crucial
assumption in adopting this perspective, which is also often exploited in
computer simulation studies (see \textit{e.g.} Refs.~\citenum{martin2,
smallenburg}) is that the particles follow the field {\em synchronously}.

In the present paper we explore, for a magnetic many-particle system, the {\em
link} between the collective, self-assembly behavior, on the one hand, and the
single-particle dynamics, on the other hand. Specifically, we consider a
ferrofluid subjected to a rotating in-plane field, where the ferrofluid is
modeled by a system of dipolar soft spheres (DSS). The same model has been
considered earlier in a computer simulation study by Murashov and
Patey,\cite{murashov} where the aim was to demonstrate that layering occurs not
only in systems of (super-)paramagnetic or polarizable particles (as those
considered by Martin \textit{et al.}\cite{martin2,martin6}), but also for
particles with permanent dipoles. Here we investigate the driven DSS system
both by Brownian dynamics (BD) computer simulations, which are described in
Sec.~\ref{sec:model}, and by theory. As a first main result, we present in
Sec.~\ref{sec:state} a full non-equilibrium ``phase'' diagram indicating the
domain of layer formation in the plane spanned by the frequency and strength of
the driving field at constant equilibrium thermodynamic parameters. Secondly,
to identify the role of mutual synchronization of the particles, we investigate
in Sec.~\ref{sec:rotational_dynamics} the rotational dynamics within layered
and unlayered states by analyzing suitable distribution functions. A similar
strategy has recently been proposed in a dynamic density functional study of
rod-like particles in rotating fields.\cite{towing} In
Sec.~\ref{sec:single_particle} we show that the breakdown of layering observed
at high frequencies in the ferrofluid system can be described by a
single-particle theory similar to those used for field-driven single
nanoparticles in viscous media.\cite{shelton} Finally, in Sec.~\ref{sec:dft},
we propose a simple equilibrium density functional approach to investigate the
role of {\em translational} entropy for layering in synchronized ferrofluid
systems. The results are in good agreement with corresponding BD simulations.
We close the paper with a brief summary and conclusions
(Sec.~\ref{sec:conclusion}).

\section{Model and simulation methods}
\label{sec:model}
In our simulations we model the colloidal suspension by a system of dipolar
soft spheres (DSS). The solvent is not explicitly taken into account. The DSS
pair potential between two spheres is comprised of a repulsive potential
$U^{\mathrm{rep}}$ and a point dipole-dipole interaction potential
\begin{multline}
    \label{eq:interaction}
    U^{\mathrm{DSS}}(\fat{r}_{ij}, \fatsym{\mu}_i, \fatsym{\mu}_j) = \\
    U^{\mathrm{rep}}(r_{ij})
    - \frac{3 (\fat{r}_{ij} \cdot \boldsymbol{\mu}_i) (\fat{r}_{ij} \cdot 
    \boldsymbol{\mu}_j)}{r_{ij}^5} + \frac{\boldsymbol{\mu}_i \cdot 
    \boldsymbol{\mu}_j}{r_{ij}^3}  .
\end{multline}
In eqn (\ref{eq:interaction}), $\fat{r}_{ij}$ is the vector between the
positions of the particles $i$ and $j$, $r_{ij}$ its absolute value, and
$\boldsymbol{\mu}_i$ is the dipole moment of the $i$th particle. The potential
$U^{\mathrm{rep}}$ is the shifted soft sphere potential, which is given by
\begin{equation}
    U^{\mathrm{rep}}(r) = U^{\mathrm{SS}}(r) - U^{\mathrm{SS}}(r_c) 
    + (r_c - r) \frac{\D U^{\mathrm{SS}}}{\D r}(r_c),
\end{equation}
where
\begin{equation}
    U^{\mathrm{SS}}(r) = 4 \epsilon \left( \frac{\sigma}{r_{ij}} \right)^{12}
\end{equation}
is the unshifted soft sphere (SS) potential for particles of radius $\sigma$.
Further, $r_c = 2.5 \sigma$ is the radius at which we cut off the potential
$U^{\mathrm{rep}}$.

We investigate the system using non-overdamped Brownian dynamics (BD)
simulations (sometimes called Langevin dynamics simulations). The corresponding
equations of motion for particles of mass $m$ and moment of inertia $I$
are\cite{murashov} 
\begin{align}
    \label{eq:eom_trans}
    m \ddot{\fat{r}}_i & = \fat{F}_i^\mathrm{DSS}
    - \xi_T \dot{\fat{r}}_i + \fat{F}^\mathrm{G}_i \\
    \label{eq:eom_rot}
    I \dot{\fatsym{\omega}}_i & = \fat{T}_i^{\mathrm{DSS} \bot}
    + \fat{T}_i^{\mathrm{ext} \bot} - \xi_R \fatsym{\omega}_i
    + \fat{T}^{\mathrm{G} \bot}_i .
\end{align}
In these equations $\xi_T= k_B T/D$ and $\xi_R = k_B T/D_r$ are friction
coefficients with $k_B$ and $T$ being Boltzmann's constant and temperature,
respectively, while $D$ and $D_r$ are the translational and rotational
diffusion constants. Furthermore $\fatsym{\omega}_i$ is the angular velocity of
particle $i$, $\fat{F}^G_i$ and $\fat{T}^G_i$ are random Gaussian forces and
torques, $\fat{T}_i^\mathrm{ext}$ are torques due to an external field, and
$\fat{T}^\bot_i = \fat{T}_i - (\fatsym{\mu}_i \cdot \fat{T}_i)
\fatsym{\mu}_i/\mu_i^2$. Their cartesian components $(\alpha, \beta = x, y, z)$
satisfy
\begin{align}
    \label{eq:bd_mean_F}
    \langle F^G_{i \alpha} (t) \rangle = 0 \\
    \langle T^G_{i \beta} (t) \rangle = 0
\end{align}
as well as
\begin{align}
    \langle F^G_{i \alpha}(t) F^G_{j \beta}(t') \rangle
    & = 6 k_B T \xi_T \delta_{ij} \delta_{\alpha \beta} \delta(t - t') \\
    \label{eq:variance_T}
    \langle T^G_{i \alpha}(t) T^G_{j \beta}(t') \rangle
    & = 6 k_B T \xi_R \delta_{ij} \delta_{\alpha \beta} \delta(t - t') .
\end{align}
As eqns (\ref{eq:bd_mean_F})-(\ref{eq:variance_T}) show, the friction
coefficients and the probability distributions of the random forces and torques
are related via the fluctuation-dissipation theorem. This ensures that the
system approaches a canonical distribution of states characterized by a
constant temperature $T$ in the absence of an external drive. To deal with the
long-ranged dipolar interactions, we used the Ewald summation method with
conducting boundaries.\cite{klapp_schoen} We have parallelized the evaluation
of the Ewald sum in our simulation with OpenMP and MPI. The equations of motion
were integrated with a Leapfrog algorithm.\cite{allentil}

The external field that the particles interact with rotates with frequency
$\omega_0$ in the $x-y$-plane and is given by
\begin{equation}
\fat{B}(t) = B_0 ( \fat{e}_x \cos \omega_0 t + \fat{e}_y \sin \omega_0 t ) .
\end{equation}

For convenience, we make use of the following reduced units: Field strength
$B_0^* = (\sigma^3/\epsilon)^{1/2} B_0$; angular frequency $\omega^*_0 = (m
\sigma^2/\epsilon)^{1/2} \omega_0$; density $\rho^* = \sigma^3 \rho$; dipole
moment $\mu^* = (\epsilon \sigma^3)^{-1/2} \mu$; and moment of inertia $I^* =
(m\sigma^2)^{-1} I$. Unless stated otherwise, the simulations were carried out
with 864 particles at density $\rho^* = 0.1$, dipole moment $\mu^* = 3$, moment
of inertia $I^* = 0.025$, and temperature $T^* = k_B T/\epsilon = 1.35$. To
verify our results, we also ran simulations with up to 4000 particles. The
translational and rotational diffusion constant were chosen to be $D = 0.1
\cdot (\epsilon \sigma^2 / m)^{1/2}$ and $D_r = 3 \cdot (m \sigma^2 /
\epsilon)^{-1/2}$, respectively, and we used a timestep of $\Delta t = 0.0025
\cdot (m \sigma^2/\epsilon)^{1/2}$. These values are consistent with those
chosen in earlier BD studies of rotating dipolar systems.\cite{murashov} We
note, however, that the effects reported in the present paper also appear for
other values of $D$ and $D_r$.

\section{Results and discussion}
\subsection{Zero field system}
\begin{figure}[ht]
    \centering
    \includegraphics[width=83mm]{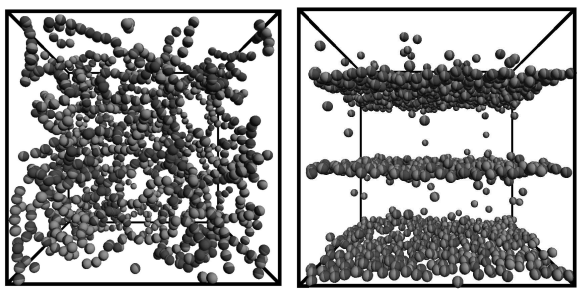}
    \caption{(a) Snapshot of the system in zero field at $\rho^* = 0.1$, $T^* =
    1.35$, and $\mu^* = 3$. (b) Snapshot of a system in a layered state. The
    strength and frequency of the field are $B^*_0 = 12$ and $\omega_0^* = 15$,
    respectively.}
    \label{fig:snaps}
\end{figure}

The zero field system, which represents our starting point, is characterized by
a large dipolar coupling strength $\lambda = \mu^2/(k_B T \sigma^3) \approx
6.7$ and a relatively low density. As expected for such a strongly coupled
system, the particles self-assemble into chainlike structures.\cite{Weis1993,
Jordanovic2011} This can be seen in the snapshot depicted in Fig.
\ref{fig:snaps}a. Our reason for considering a system of a coupling strength
this high is that this seems to be a prerequisite for layer formation. Indeed,
irrespective of the field strength, we did not observe any layering for values
of $\lambda$ that are smaller than approximately $4.6$ (at the temperature $T^*
= 1.35$).

Contrary to $\lambda$, our choice of the density is less restricted, since the
layering phenomenon persists over a wide range of densities (at least up to
$\rho^* = 0.4$). However, choosing the small density of $\rho^* = 0.1$ has the
advantage that layers are easily discernible.

\subsection{The layering effect}
\label{sec:state}
We now consider the same system in rotating fields of various strengths $B_0^*$
and frequencies $\omega_0^*$. For sufficiently large $B_0^*$ and not too high
frequencies (see below), the particles arrange themselves into layers. An
example of this is shown in Fig. \ref{fig:snaps}b. This phenomenon was first
explained by Halsey, Anderson, and Martin.\cite{martin1} They realized that the
time-averaged potential between two particles $i$ and $j$ that rotate with the
same angular frequency (given by the external field) and are aligned with each
other, \textit{i.e.} rotate circularly in a synchronized fashion with
\begin{equation}
    \fatsym{\mu}_i (t) = \fatsym{\mu}_j (t)
    = \mu ( \fat{e}_x \cos \omega_0 t + \fat{e}_y \sin \omega_0 t ),
\end{equation}
is given by
\begin{multline}
    \label{eq:dipole_avg}
    U^{\mathrm{ID}} (\fat{r}_{ij}) = \tau^{-1} \int_{t_0}^{t_0 + \tau}
    U^\mathrm{D}(\fat{r}_{ij}, \fatsym{\mu}_i(t), \fatsym{\mu}_j(t))
    \D t \\
    = - \mu^2 \frac{(1 - 3 \cos^2 \Theta_{ij})}{2 r^3_{ij}} .
\end{multline}
In this equation, $U^\mathrm{D}$ is the dipole-dipole potential (see eqn
(\ref{eq:interaction}), $\tau = 2 \pi / \omega_0$ is the oscillation period,
and $\Theta_{ij}$ is the angle between the interparticle vector $\fat{r}_{ij}$
and the direction perpendicular to the plane of the field. As shown by the last
line in eqn (\ref{eq:dipole_avg}), the time-averaged potential corresponds to
an inverted dipolar (ID) potential, which is attractive if the angle
$\Theta_{ij}$ satisfies $\cos^2 \Theta_{ij} < 1/3$, \textit{i.e.} if the
particles $i$ and $j$ are approximately in the same plane with respect to the
field. Conversely, if the angle $\Theta_{ij}$ satisfies $1/3 < \cos^2
\Theta_{ij}$, the particles repel each other. This direction dependence of the
ID potential explains why layers are a favorable configuration for a driven
system in which the particles rotate synchronously.

Note that for the above argument to hold, the translational motion of the
particles should be small compared to their rotational motion. More precisely,
during one rotational period they should migrate much less than their own
diameter.\cite{martin1}

\begin{figure}[ht]
    \centering
    \includegraphics[width=30mm]{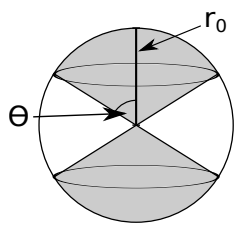}
    \caption{Sketch of the polar and equatorial regions used in the definition
    of the order parameter.}
    \label{fig:order_parameter_sketch}
\end{figure}

In the following we aim to determine more precisely the range of frequencies
and field strengths at which layering occurs. To do that, we need a suitably
defined order parameter. We tested several ones and compared them with one
another. The order parameter that we will use here is given by
\begin{equation}
    \label{eq:order_parameter}
    \psi = \frac{1}{N} \sum_{i=1}^{N} \langle n_i \rangle,
\end{equation}
where $N$ is the total number of particles, $\langle \cdots \rangle$ denotes a
time-average, and $n_i$ is defined as follows: Consider a sphere of radius
$r_0$ around particle $i$. Divide that sphere into two parts, one of which is
given by the points within the sphere whose distance vector to particle $i$
together with the $z$-axis encloses an angle $\Theta$ satisfying $-0.5 < \cos
\Theta < 0.5$ (see Fig. \ref{fig:order_parameter_sketch}). If there are more
(less) particles in this equatorial volume than in the polar volume around
particle $i$, set $n_i = 1$ ($-1$); if there are the same number of particles,
set $n_i = 0$.
Note that the radius $r_0$ was set to $8 \sigma$. Smaller as well as larger
radii $r_0$ decrease the performance of the order parameter as we found by
comparing the order parameter with the actual order observed in the system. 

\begin{figure}[ht]
    \centering
    \includegraphics[width=80mm]{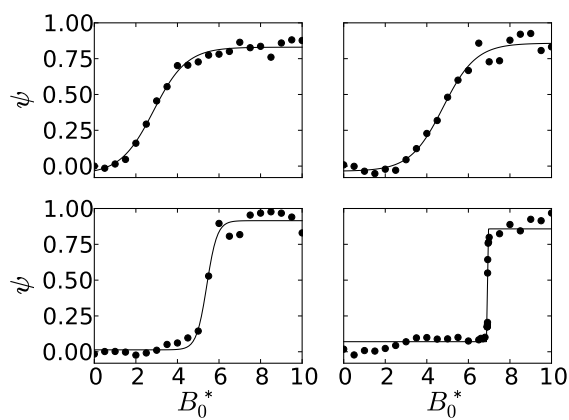}
    \caption{The order parameter $\psi$ at constant angular frequency (a)
    $\omega_0^* =1$, (b) $20$, (c) $30$, (d) $40$.}
    \label{fig:order_parameter_bd}
\end{figure}

Representative examples for the behavior of the resulting order parameter at
constant angular frequency but increasing field strength are given in Fig.
\ref{fig:order_parameter_bd}. As can be seen, in all the cases the value of
$\psi$ grows with the field strength until it almost reaches a value of $1$.
Since the layers are usually not perfectly defined in our Brownian dynamics
simulations, the order parameter typically takes on values that are slightly
smaller than $1$ even at very high field strengths.

One also finds from Fig. \ref{fig:order_parameter_bd} that there is a
qualitative difference in the behavior of $\psi$ at high and low frequencies:
The order parameter grows much more steeply at large frequencies, which means
that the layers do not slowly emerge upon increasing the field strength but
appear very rapidly.

By inspecting snapshots corresponding to a given value of the order parameter,
it turned out that the value $\psi_0 \approx 0.6$ may serve as an (approximate)
lower limit for layer formation.

\begin{figure}[ht]
    \centering
    \includegraphics[width=83mm]{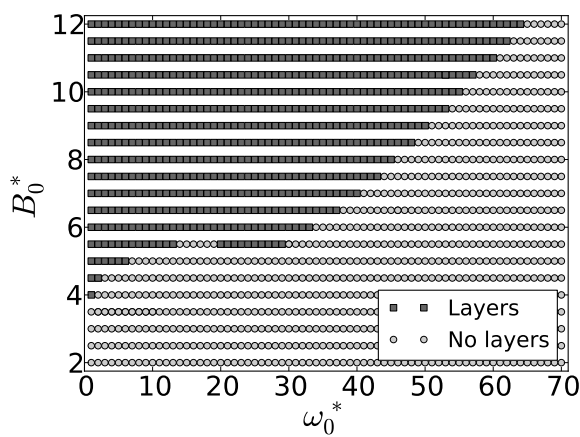}
    \caption{Occurrence of layers depending on field strength and frequency of
    the driving field. The system parameters are chosen as described in
    Sec.~\ref{sec:model}.}
    \label{fig:wEbd}
\end{figure}

Based on that criterion, we have scanned a broad range of frequencies and field
strengths for the occurrence of layers. The results of this exploration of the
parameter space are summarized in Fig. \ref{fig:wEbd}. Note that every
simulation was started from a random configuration to avoid
hysteresis-like effects.

Within the layered ``state'', the translational structure of one layer is
disordered and becomes more and more homogeneous with larger $\omega_0^*$. In
particular there is no pronounced hexagonal order as observed in earlier
studies,\cite{martin2} even though the particles tend to have six nearest
neighbors at high $\omega_0$. This absence of pronounced in-plane order is
probably a consequence of the low density considered ($\rho^* = 0.1$) and the
Brownian random forces. Furthermore, depending on the initial conditions, we
typically observe two or three layers in our simulation box ($N = 864$), which
corresponds to an average vertical distance between the layers of about seven
to ten particle diameters.

The figure shows that the $\omega_0^*-B_0^*$ diagram is separated into a
layered and a non-layered region. Upon increasing the frequency from zero, the
boundary first remains at roughly constant field strength, until it begins to
rise with the frequency. This behavior is mirrored in Fig.
\ref{fig:order_parameter_bd}. The larger the frequency, the higher the field
strength at which the order parameter attains large values.

\begin{figure}[ht]
    \centering
    \includegraphics[width=83mm]{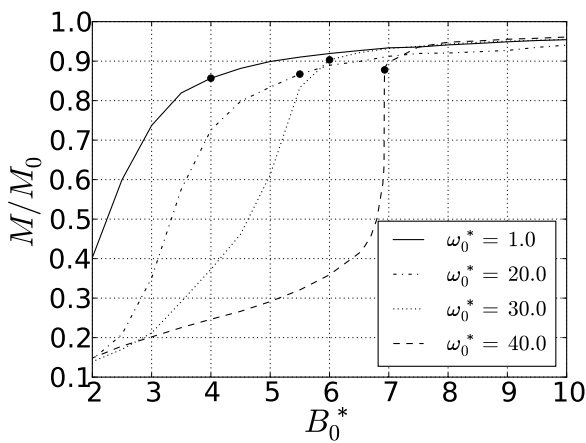}
    \caption{Absolute value of the magnetization normalized with respect to its
    saturation value over field strength at different rotational frequencies.
    The dots indicate after which point the system is considered layered
    according to our order parameter.}
    \label{fig:magnetization}
\end{figure}

A similar picture emerges from Fig. \ref{fig:magnetization}, where we have
plotted the normalized absolute value of the magnetization, \textit{i.e.}
$M/M_0 = \langle | \fat{M}(t) | \rangle/M_0$ with $\fat{M}(t) = \sum_{i=1}^{N}
\fatsym{\mu}_i(t)$ and $M_0 = N \mu$, as a function of $B_0^*$ for several
frequencies. Note that $|\fat{M}(t)|$ is essentially independent of time for
the states considered in Fig. \ref{fig:magnetization}. Clearly, the
magnetization behaves differently in the regimes of small and large
frequencies. One also finds from Fig. \ref{fig:magnetization} that a degree of
magnetization of more than $80$ percent is required for layer formation to
occur.

In the following subsections, we will discuss the emergence and breakdown of
layering in the different frequency regimes in more detail.
Before doing so, it is worth to briefly comment on a technical issue
encountered in our exploration of the parameter space (see Fig. \ref{fig:wEbd})
that concerns the behavior of the rotational temperature $T_{\mathrm{rot}} =
1/(2(N-1)) \sum_{i=1}^N I \omega_i^2$. Upon increasing the driving frequency
$\omega_0^*$ from zero (at fixed $B_0^*$), we typically also find
$T_{\mathrm{rot}}$ to increase, while its translational counterpart
$T_{\mathrm{trans}} = 1/(3(N-1)) \sum_{i=1}^N m \fat{v}_i^2$ stays
approximately constant (close to the input value $T$). Similar temperature
drifts have been observed in other non-equilibrium systems such as fluids in
shear flow. In the latter context, the temperature is often redefined with
respect to the differences between the actual velocity of the particle and that
of the flow field.\cite{Delhommelle2004} Using a similar definition here
(involving the difference between $\omega_i$ and $\omega_0$), we find that this
temperature is still not equal to $T$, but remains essentially constant over a
broad range of frequencies. We also note that both the temperature drift and
the actual location of the layer boundary in the $\omega_0^*-B_0^*$ diagram
depend on the chosen value of the rotational friction constant.

\subsection{Rotational dynamics in the layered state}
\label{sec:rotational_dynamics}
As mentioned earlier, the key argument for the appearance of the layers is that
the time-averaged interaction between two {\em fully synchronized} rotating
dipoles favors an in-plane configuration. In the following, we will investigate
in more detail to what extent this assumption is actually fulfilled within the
layered region indicated in Fig. \ref{fig:wEbd}. To this end, we consider the
distribution $f$ of the phase differences $\phi_i$ between the dipolar vector
of particle $i$ in the $x-y$-plane and the external field. More precisely, we
define $f$ as
\begin{multline}
    \label{eq:dist_phi}
    f(\phi) = \frac{1}{N \Delta \phi} \Bigg< \sum_{i=1}^N \Theta(\phi_i
    - n \Delta \phi) \\
    - \Theta(\phi_i - (n+1) \Delta \phi) \Bigg>,
\end{multline}
where $\Theta$ is the Heaviside function, $\Delta \phi$ is the interval length
to which we want to resolve the distribution, $n$ is a positive integer or zero
that satisfies $n \Delta \phi \leq \phi < (n+1) \Delta \phi$, and, as before,
$\langle \cdots \rangle$ denotes a time-average.

\begin{figure}[ht]
    \centering
    \includegraphics[width=83mm]{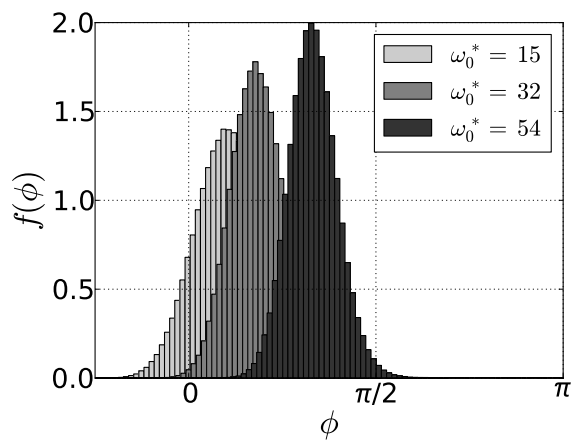}
    \caption{Distribution of the phase differences at $B_0^* = 10$ and three
    frequencies $\omega_0^*$ with $\Delta \phi = \pi / 48$. The systems are in
    layered states (see Fig. \ref{fig:wEbd}).}
    \label{fig:dist_multi_phi}
\end{figure}

We start by considering systems that are driven by fields of considerable
strength ($B_0^* = 10$) with frequencies that admit layer formation
(\textit{cf.} Fig. \ref{fig:wEbd}). Results for the distribution $f$ at three
such frequencies $\omega_0^*$ are given in Fig. \ref{fig:dist_multi_phi}. For
each value of $\omega_0^*$ one observes a single, pronounced peak, reflecting a
synchronized ``state'', in which the particles follow the field at constant
phase difference.
Note that the larger $\omega_0^*$, the larger the phase difference between the
particles and the field. This is not too surprising since an increase in the
driving frequency implies an increase in the rotational friction due to the
(implicit) solvent and the presence of neighboring particles. Further note that
even though eqn (\ref{eq:dist_phi}) contains a time-average, the phase
distributions of these layered systems are essentially independent of time.

To investigate the degree to which the particles actually rotate in the plane
of the field, we also consider the distribution of the $z$-components of the
angular frequencies
\begin{equation*}
    g(\omega_z^*) = \frac{1}{N \Delta \omega}
    \Bigg< \sum_{i=1}^N \Theta(\omega_{i,z}^*  - n \Delta \omega) \\
    - \Theta(\omega_{i,z}^* - (n+1) \Delta \omega) \Bigg> .
\end{equation*}

\begin{figure}[ht]
    \centering
    \includegraphics[width=83mm]{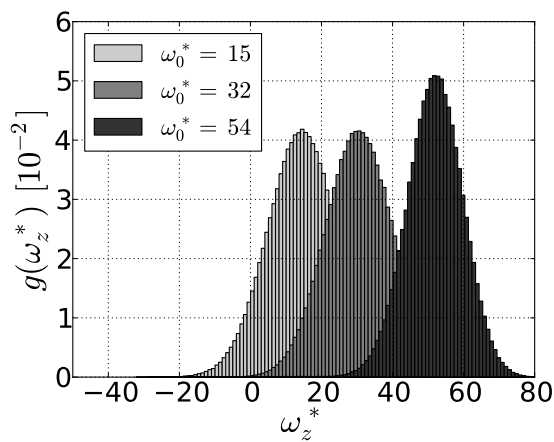}
    \caption{Distributions of the $z$-component of the angular frequencies.
    Parameters are as in Fig. \ref{fig:dist_multi_phi}. The interval $\Delta
    \omega$ is set to $1$.}
    \label{fig:dist_multi_omega}
\end{figure}
In an ideal situation, in which the dipoles rotate perfectly with the field,
the distribution $g$ would have a single, sharp peak at $\omega_z^* =
\omega_0^*$. Simulation results for $g$ in the true many-particle system are
shown in Fig. \ref{fig:dist_multi_omega}, where we have picked out the
``states'' already considered in Fig. \ref{fig:dist_multi_phi}. As expected in
the layered regime, the functions $g$ are characterized by one central peak
around $\omega_z^* \approx \omega_0^*$. However, we also see that there is a
significant broadness in the distribution (as there is in the corresponding
peaks of $f$).

\begin{figure}[ht]
    \centering
    \includegraphics[width=83mm]{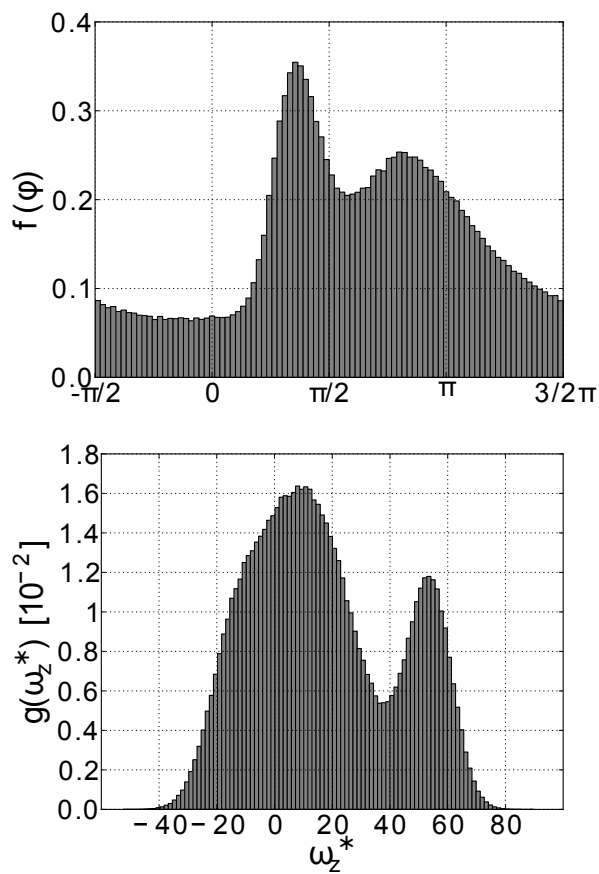}
    \caption{(a) Distribution of the phase differences of the system at $B_0^*
    = 10$ and $\omega_0^* = 55.8$, \textit{i.e.} just outside of the region of
    layer formation. The resolution is $\Delta \phi = \pi/40$. (b) Distribution
    of $\omega^*_z$ for a system at $B_0^* = 10$ and $\omega_0^* = 55.8$ with
    $\Delta \omega = 1$.}
    \label{fig:dists_55_8}
\end{figure}

\begin{figure}[ht]
    \centering
    \includegraphics[width=83mm]{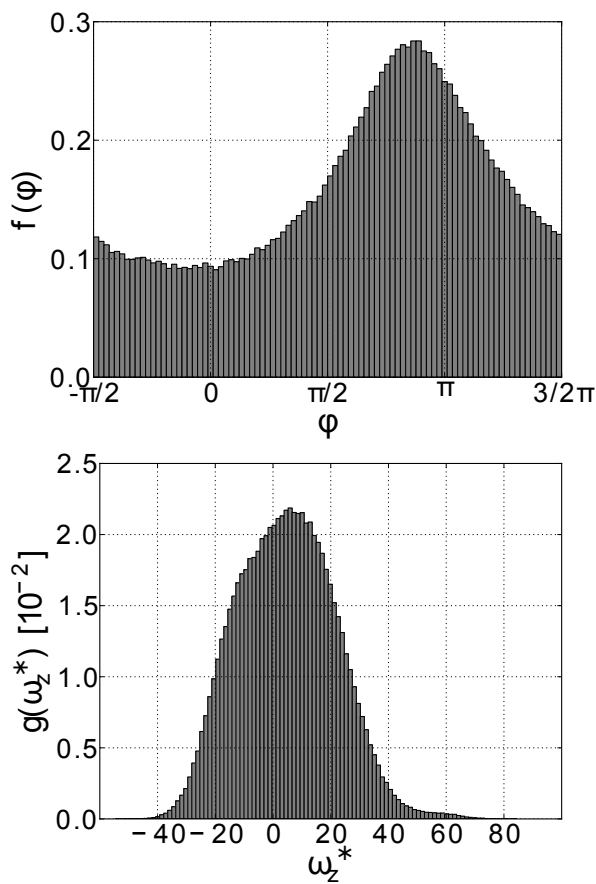}
    \caption{(a) Distribution of phase differences of the system at $B_0^* =
    10$ and $\omega_0^* = 60$. The system is unlayered. (b) Distribution of
    $\omega^*_z$ of the system at $B_0^* = 10$ and $\omega_0^* = 60$. The
    system is unlayered.}
    \label{fig:dists_60}
\end{figure}

Finally, above a certain frequency, the layers disappear. This is reflected in
the emergence of a double-peaked structure in the distribution of the phase
differences, as illustrated in Fig. \ref{fig:dists_55_8}a. Moreover, we found
that the non-averaged distribution of the phase differences is not independent
of the time anymore. However, since we could not identify any {\em systematic}
time-dependence in this regime, we restrict ourselves to considering the
averaged distribution. The first peak in $f$ at $\phi \approx \pi/4$ is due to
particles that can still temporarily follow the field, whereas particles that
are not able to do so anymore cause the structure of the rest of the
distribution. The breakdown of layering is also visible in the distribution
$g$. Contrary to what is seen in a layered system, the angular frequencies of
the majority of the particles are distributed around $\omega^*_z \approx 0$ as
shown in Fig. \ref{fig:dists_55_8}b. The much smaller peak at approximately the
frequency of the external drive shows that only a small fraction of the
particles follow the field at any given time. This fraction is further
decreased as the frequency $\omega_0^*$ of the driving field increases. Typical
distributions $f$ and $g$ at values of $\omega_0^*$ outside the layered regime
are shown in Figs. \ref{fig:dists_60}a and \ref{fig:dists_60}b, respectively.
Note that the roughly symmetric distribution of $\omega_z^*$ around
approximately zero in Fig. \ref{fig:dists_60}b indicates that the particles are
as likely to rotate in the direction of the field as they are to rotate in the
opposite direction.


Further note that at the large values of $B_0^*$ considered in this section,
the transition between states with the particles following the field at fixed
phase difference and states where this is not true anymore happens in a very
small range of frequencies.

\subsubsection{Effective single-particle theory.~~}
\label{sec:single_particle}
To understand the character of the high-frequency boundary between layered and
non-layered states in more detail, we now aim to construct an effective theory
that describes a single dipolar particle rotating in a viscous medium. A
similar consideration has been suggested for optically torqued nanorods by
Shelton \textit{et al.}.\cite{shelton} Clearly, such a single-particle approach
cannot grant us direct insight into the formation of layers. However, it may
help us to improve our understanding of the rotational dynamics isolated from
many-particle effects. For simplicity, we assume that the rotational motion of
the particle is restricted to the plane of the field and that it experiences
rotational friction with friction constant $\gamma$. Then the rotational
equation of motion is
\begin{equation}
\label{eq:eom_alpha}
I \ddot{\alpha} = - \gamma \dot{\alpha} + \mu B_0 \sin(\omega_0 t - \alpha),
\end{equation}
where $\alpha$ is the angle between the dipolar orientation and an arbitrary
axis within the plane of the field. Equation (\ref{eq:eom_alpha}) can be
rewritten in terms of the phase difference $\phi = \omega_0 t - \alpha$ as
\begin{equation}
    \label{eq:eom_phi}
    I \ddot{\phi} + \gamma \dot{\phi} = \gamma \omega_0 - \mu B_0 \sin \phi .
\end{equation}
We first consider the simplified case of negligible moments of inertia,
\textit{i.e.} an overdamped situation. Then eqn (\ref{eq:eom_phi}) reduces to
the first order equation
\begin{equation}
\label{eq:non_linear}
\frac{\D \phi}{\D \tau} = \frac{\omega_0}{\omega_c} - \sin \phi ,
\end{equation}
where $\omega_c = \mu B_0/\gamma$ and $\tau = \omega_c t$. This nonlinear
differential equation appears in various contexts such as the description of
overdamped pendula, superconducting Josephson junctions, and the synchronized
emission of light by fireflies.\cite{strogatz,shelton} For $0 \leq \omega_0 <
\omega_c$ it has two fixed points characterized by $\dot{\phi} = 0$
(\textit{i.e.} constant phase difference): One solution is a global attractor
with $\phi = \arcsin (\omega_0/\omega_c)$, and the other one is unstable with
$\phi = \pi - \arcsin (\omega_0/\omega_c)$. These two solutions correspond to
the phase differences at which the torque due to friction equals the torque
that is due to the field. At $\omega_0 = \omega_c$, \textit{i.e.} at $\phi =
\pi/2$, the two solutions form a saddle-node bifurcation and there are no fixed
points for $\omega_0 > \omega_c$. At these high frequencies, the maximal torque
that can be exerted by the field is insufficient to balance the frictional
torque. The solution emerging after the bifurcation is a limit cycle with
$\dot{\phi} > 0$.

\begin{figure}[ht]
    \centering
    \includegraphics[width=83mm]{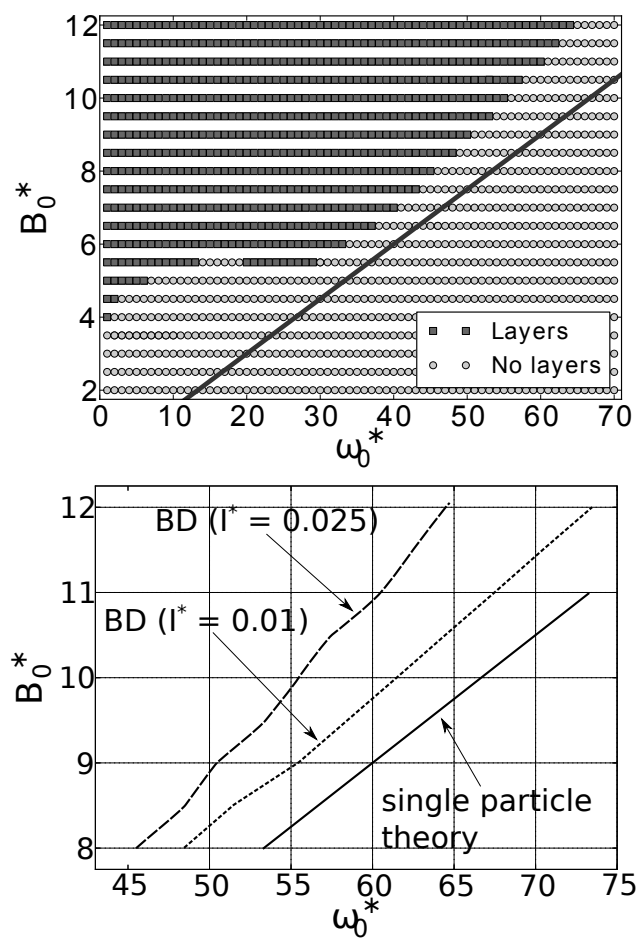}
    \caption{(a) The solid line shows the critical frequencies $\omega_c = \mu
    B_0/\xi_R$ that are predicted by the single-particle theory in the BD
    frequency-field strength diagram (see Fig. \ref{fig:wEbd}). (b) Influence
    of the moment of inertia on the end of layer formation (dashed line: $I^* =
    0.025$, dotted line: $I^* = 0.01$). The solid line indicates the
    frequencies $\omega_c^*$.}
    \label{fig:crit}
\end{figure}
To which extent does the single-particle approach describe the true
many-particle system of our BD simulations? In Fig. \ref{fig:crit}a, the
frequencies $\omega_c$ (with $\gamma \equiv \xi_R$, see eqn (\ref{eq:eom_rot}))
are plotted into the $\omega_0^*-B_0^*$-state diagram (Fig. \ref{fig:wEbd}). At
large frequencies $\omega_0^*$ and field strengths $B_0^*$, the straight line
representing $\omega_c$ has a slope similar to that of the boundary of the
layered regime. This supports the idea that it is the (rotational) friction
which eventually yields a breakdown of the synchronous rotations, and thus, the
layering. A further observation from Fig. \ref{fig:crit}a is that the true
boundary frequencies (at given $B_0^*$) are somewhat smaller than $\omega_c$.
One seemingly obvious reason for these deviations is that the effective theory
neglects any many-particle effects. Moreover, it does not take the Brownian
random contributions into account that mimic the solvent ``kicks'' in eqn
(\ref{eq:eom_rot}). Both these factors could introduce perturbations of the
effective field that acts on a particle. Thereby the synchronized state could
be destabilized already at frequencies $\omega < \omega_c$. However, as it
turns out, the more significant reason for the premature stop of layering is
that the BD equations of motion involve (rotational) inertial terms, which are
neglected in our single-particle approach.

To check this point, we have performed additional BD simulations with a lower
moment of inertia ($I^* = 0.01$). The resulting frequencies characterizing the
boundary of the layered state are shown in Fig. \ref{fig:crit}b along with the
original result ($I^* = 0.025$) and the line $\omega_c$. Clearly, decreasing
the moment of inertia moves the true boundary substantially closer to the
single-particle result.

Finally, we note that the influence of the inertial (rotational) term can also
be captured within our effective single particle theory. For $I \neq 0$, eqn
(\ref{eq:eom_phi}) can be written as
\begin{equation}
    \label{eq:non_linear_inertial}
    \frac{\D^2 \phi}{\D \tau'^2} + \nu \frac{\D \phi}{\D \tau'}
    = \frac{\omega_0}{\omega_c} - \sin \phi
\end{equation}
with $\nu = \gamma / \sqrt{\mu B_0 I}$ and $\tau' = \sqrt{\mu B_0 / I} t$.
Similar to (\ref{eq:eom_phi}), this differential equation has a bifurcation at
$\omega_c$,\cite{strogatz} which means that the location of the line $\omega_c$
in Fig. \ref{fig:crit}a remains unchanged.\cite{strogatz} As before, the only
stable solution at driving frequencies that are larger than $\omega_c$ is a
limit cycle. But additionally, for sufficiently small $\nu$, it has a second
bifurcation for some $\omega'$ with $\omega' < \omega_c$ as shown by Argentina
\textit{et al.} while investigating the transition between annihilation and
preservation of colliding waves.\cite{argentina} This second bifurcation
introduces a regime in which the limit cycle can coexist with the stable
rotation. From the perspective of a many-particle system, one may speculate
that the presence of the second solution perturbs the rotation with constant
phase difference (\textit{i.e.}, $\dot{\phi} = 0$).

\subsection{A density functional approach to layering in a perfectly synchronized system}
\label{sec:dft}
We now consider systems at relatively low driving frequencies
($\omega_0^{*}\lesssim 30$), where, for sufficiently large field strengths
$B_0^{*}$, the dipole vectors can follow the field in a perfectly synchronized
fashion (see the discussion in the preceding section). According to our
``phase'' diagram in Fig. \ref{fig:wEbd}, the field strength required to induce
such synchronous and, at the same time, layered states, is about
$B_0^{*}\approx 4-6$ for $\omega_0^{*}\lesssim 30$. The corresponding
dipole-field coupling parameter $\mu B_0/k_{\mathrm{B}}T= \mu^*B_0^*/T^*
\approx 12$ is significantly larger than the dipole-dipole coupling parameter
($\lambda \approx 6.7$). Nevertheless, as seen in Figs.
\ref{fig:order_parameter_bd}a and b as well as Fig. \ref{fig:magnetization},
increasing $B_0^{*}$ from zero at low driving frequencies yields a rather slow
increase of the order parameter $\psi$ and the magnetization amplitude.

Given the apparent interconnectedness between the rotational dynamics of the
individual dipoles and the layering of the particles, we ask in the present
section whether synchronization leads {\em automatically} to layering. Indeed,
even in a perfectly rotating system, one would expect that the spatial
symmetry-breaking associated with layering yields a decrease of translational
entropy and thus may be unfavorable.

To investigate this question we employ {\em equilibrium} density functional
theory (DFT) for a system in which the dipole rotations are perfectly
synchronized. Under such conditions the particles effectively interact via the
time-averaged (inverted) dipolar potential given in eqn (\ref{eq:dipole_avg}).
By using this potential, the problem thus reduces to searching for an
equilibrium {\em phase transition} in a system with effectively static
interactions.

Our density functional approach is based on the perturbation expansion of the
free energy originally proposed by Ramakrishnan and Yussouff in the context of
fluid-solid transitions.\cite{ramakrishnan} Up to second order in the density,
the difference between the Helmholtz free energy of a volume $V$ of a system
with non-uniform density $\rho(\fat{r})$ and a reference system with
homogeneous density $\rho_0$ is given by\cite{haymet}
\begin{multline}
    \label{eq:df_pert}
    \frac{\Delta \mathcal{F}}{V} = \frac{1}{\beta V} \int_V \D^3r [\log(\Lambda^3
    \rho(\fat{r})) - 1] \\
    - \frac{1}{\beta V} \int_V \D^3r \rho_0 [ \log(\lambda^3 \rho_0) - 1] \\
    - \frac{1}{2 \beta V} \int_V \D^3r_1 \int_{\real^3} \D^3r_2 c(\fat{r}_1 -
    \fat{r}_2)|_{\rho_0} \Delta \rho(\fat{r}_1) \Delta \rho(\fat{r}_2).
\end{multline}
In eqn (\ref{eq:df_pert}), $\Delta \rho(\fat{r}) = \rho(\fat{r}) - \rho_0$ with
$\int_V \D^3 r \Delta \rho(\fat{r}) = 0$, $\Lambda$ is the thermal wavelength,
and $c(\fat{r})|_{\rho_0}$ is the direct correlation function of the
homogeneous system.

Here we employ the random phase approximation (RPA) to calculate the direct
correlation function.\cite{hansen_mcdonald} Assuming a hard sphere interaction
in addition to the inverse dipolar potential $U^{\mathrm{ID}}$ (eqn
(\ref{eq:dipole_avg})), the RPA amounts to setting
\begin{equation}
    \label{eq:c}
    c(\fat{r}) = 
    \begin{cases}
        c^{\mathrm{PY}}(r),  & r \leq \sigma \\
        -\beta U^{\mathrm{ID}}(\fat{r}), & r > \sigma ,
    \end{cases}
\end{equation}
where we used the Percus-Yevick direct correlation function,
$c^{\mathrm{PY}}$,\cite{hansen_mcdonald} for the hard-sphere part. Note that
within the RPA, the effects of the contributions of the long-ranged inverse
dipolar interaction are treated in a mean-field fashion. To check this point,
we have also calculated $c(\fat{r})$ numerically by solving the mean-spherical
(MSA) integral equations.\cite{hansen_mcdonald} However, the changes in the
free energies were found to be marginal.

As a simple ansatz for the density profile in the layered state, we use
\begin{equation}
    \label{eq:rho_cos_ansatz}
    \rho(\fat{r}) = \rho(z) = \rho_0 + \tilde{\rho} \cos(kz).
\end{equation}
Inserting this ansatz into eqn (\ref{eq:df_pert}), we find
\begin{equation}
    \label{eq:df_pert_eval}
    \frac{\Delta \mathcal{F}}{A}  = \frac{1}{\beta} \int_0^{\lambda_L} \D z \rho(z)
    \log \left( \frac{\rho(z)}{\rho_0} \right)
    - \lambda_L \frac{\tilde{\rho}^2}{4 \beta} \tilde{c}(k),
\end{equation}
where $\Delta \F$ is the free energy of the volume $A \cdot \lambda_L$, $A$ is
an area in the $x-y$-direction and $\lambda_L = 2 \pi / k$. Further,
$\tilde{c}$ is the Fourier transform of $c$ and $\tilde{c}(k) \equiv
\tilde{c}(k \fat{e}_z)$. In the RPA, we have
\begin{equation}
    \label{eq:ck}
    \tilde{c}(k) = 4 \pi \left( \int_0^\sigma \D r r^2 j_0(kr) c^{\mathrm{PY}}(r)
    + \mu^2 \beta \frac{j_1(k \sigma)}{k \sigma} \right),
\end{equation}
where $j_n$ are spherical Bessel functions of order $n$. (For the treatment of
the dipolar interactions in eqn (\ref{eq:ck}), see Ref. \citenum{wei})
\begin{figure}[h!t]
    \centering
    \includegraphics[width=83mm]{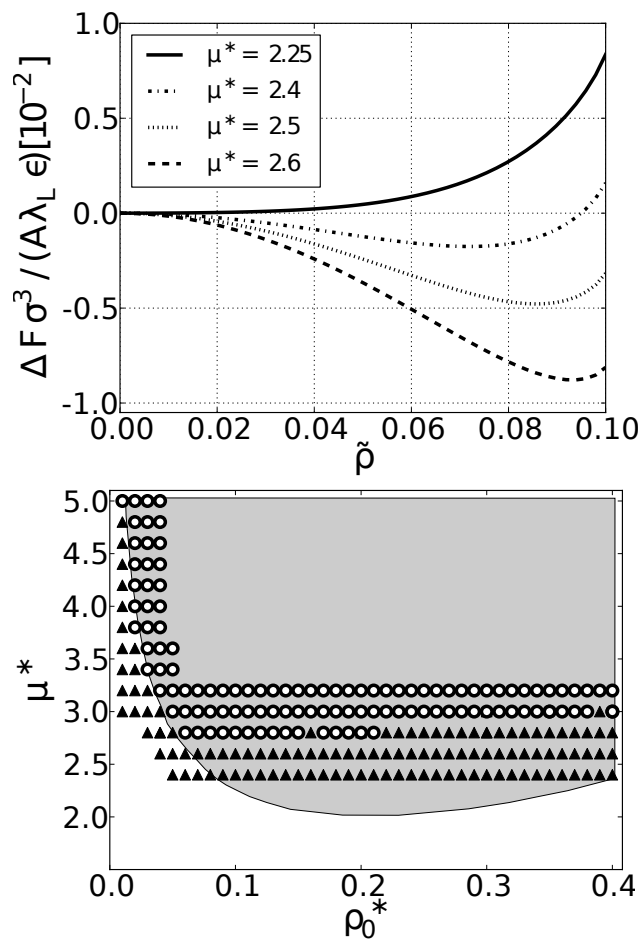}
    \caption{(a) Free energy difference between a layered and a homogeneous
    state as a function of the coefficient $\tilde{\rho}$ (see eqn
    (\ref{eq:rho_cos_ansatz})) at different values of the parameter $\mu^*$.
    The case $\tilde{\rho} = 0$ corresponds to the homogeneous  solution. The
    overall density is set to $\rho_0^* = 0.1$. (b) Equilibrium phase diagram
    of a perfectly synchronized system. The gray area indicates the stability
    range of the layered state according to our DFT calculations. Also shown
    are BD results (at $\omega_0^* = 8, B_0^* = 50$) with the open circles
    (solid triangles) representing layered (non-layered) states.}
    \label{fig:dft}
\end{figure}
We now use eqn (\ref{eq:df_pert_eval}) to search for a phase transition between
the homogeneous and the layered state. In principle, this search requires a
minimization of $\Delta \F/A$ with respect to both the parameters
$\tilde{\rho}$ and $k$ that characterize the inhomogeneity of the system (see
eqn (\ref{eq:rho_cos_ansatz})). It turns out, however, that $\Delta \F/A$
becomes minimal with respect to $k$ for $k\rightarrow 0$, which corresponds to
an {\em infinite} distance between the layers. Clearly, this is not compatible
with the implicit assumption of a finite wavelength. Therefore we have fixed
the parameter $k = 2\pi/\lambda_L$ to physically reasonable values,
\textit{i.e.} to values suggested by our BD simulations. At $\rho_0^{*}=0.1$,
we find an average layer distance of approximately $7.2 \sigma$ (see below).
This leaves the coefficient $\tilde{\rho}$ as the only minimization parameter.
Results for the function $\Delta \F(\tilde{\rho})/(A \lambda_L)$ with fixed
distance $\lambda_L = 7.2 \sigma$ between the layers at various values of the
parameter $\mu^*$ are plotted in Fig. \ref{fig:dft}a.

The different curves in Fig. \ref{fig:dft}a reveal a behavior typical of a
second-order phase transition. For $\mu^* \lesssim 2.27$, the free energy has
only one minimum at $\tilde{\rho}=0$ corresponding to an homogeneous state.
This changes at $\mu^*_{\mathrm{c}} \approx 2.27$: For larger values of
$\mu^*$, the solution at $\tilde{\rho}=0$ represents a maximum, and the only
minimum occurs for $\tilde{\rho} > 0$. The corresponding negative values of
$\Delta \F/A$ indicate that it is indeed the layered state which is now
globally stable.

We have repeated the DFT calculations for a number of densities in the range
$0.01 \leq \rho_0^* \leq 0.4$. To find reasonable values for the corresponding
wavelengths $\lambda_{\mathrm{L}}$ in the layered state, we ran BD simulations
at fixed dipole moment $\mu^* = 3.4$, frequency $\omega_0^* = 8$, and field
strength $B_0^*=50$. With this choice of the parameters, the particles are
almost perfectly aligned, justifying the key assumption of our DFT approach.
Fitting the resulting distances as functions of $\rho_0$, we found the
approximate relation $d / \sigma \approx 1.05 \rho^{*-0.84}$, which was then
used as an input in the DFT (\textit{i.e.}, $\lambda_{\mathrm{L}} = d$).

The resulting phase diagram in the $\rho_0^{*}-\mu^{*}$-plane is plotted in
Fig. \ref{fig:dft}b. It is seen that the DFT predicts a layering transition for
all but the smallest densities ($\rho_0^{*} \gtrsim 0.01$) in the shown
parameter range, with the actual values of $\mu^*_{\mathrm{c}}$ varying
substantially with $\rho_0^{*}$. Indeed, the lowest threshold is found at
$\rho_0^{*}\approx 0.2$. Also shown in Fig. \ref{fig:dft}b are BD results for
the appearance of layers in nearly perfectly synchronized systems
($\omega_0^*=8$, $B_0^*=50$) at various values of $\mu^{*}$. As in Sec.~\ref{sec:state},
the presence of layers was detected on the basis of the order parameter defined
in eqn (\ref{eq:order_parameter}), yet with a slightly different definition of
the cutoff radius entering the order parameter ($r_0 = d$). Comparing BD and
DFT, it is seen that the DFT predicts the true phase boundary in perfectly
synchronized systems in a qualitatively correct manner (including the strong
increase of $\mu^*_c$ upon $\rho_0^*\rightarrow 0$). Moreover, the DFT results
are also quite reasonable from a quantitative point of view.

From a physical perspective, clearly the most important conclusion is that even
in a perfectly synchronized system, a sufficient decrease of interaction energy
stemming from the time-averaged dipolar potential is required to overcome the
entropy cost due to layering.

Finally, we briefly discuss our DFT results in the light of a recent Monte
Carlo study by Smallenburg and Dijkstra,\cite{smallenburg} who obtained full
equilibrium phase diagrams of systems interacting with inverted dipolar
interactions. To model the short-range part of the interaction, Smallenburg and
Dijkstra used either just hard spheres or hard spheres with an additional
Yukawa repulsion.\cite{smallenburg} In the first case, layer-like structures
were only observed in the gas-liquid coexistence region. On the contrary, the
Yukawa system exhibits a stable layered phase with fluid-like in-plane
structure. Comparing these latter results to our DFT predictions, we find that
the predicted strength of the inverted dipolar interactions required for layer
formation is indeed comparable. On the other hand, we find the onset of layer
formation at much lower densities. Apart from the obvious approximations in our
theory, we also attribute these deviations to the fact that the repulsive
Yukawa interaction used in Ref.~\citenum{smallenburg} is much stronger than our
soft sphere one.

\section{Conclusions}
\label{sec:conclusion}
In this study we have combined BD computer simulations, an effective
single-particle theory, and an (equilibrium) density functional approach to
explore the dynamic behavior of systems of dipolar particles in planar rotating
fields. 

One main result from our BD simulations is a non-equilibrium ``phase'' diagram
identifying the domain of layered states in the $\omega_0$-$B_0$ plane (at
constant particle density and dipolar coupling strength). At low driving
frequencies, the change from unlayered to layered (and fully synchronized)
structures occurring upon increase of $B_0$ is related to a quasi-equilibrium
phase transition, \textit{i.e.} a many-particle phenomenon. The transition is
induced by the competition between the time-averaged, inverted dipolar
interactions favoring in-plane configurations and the loss of translational
entropy accompanying the one-dimensional translational order. While this
competition also occurs for systems of polarizable or superparamagnetic
particles, the additional complication in the present system of permanent
dipoles is that the field first needs to overcome the dipolar fluctuations.
Though we have neglected this issue in our DFT approach, we would expect that
the fluctuations just shift the transition predicted by the DFT towards larger
field strength. 

Completely different behavior is found at high frequencies and field strengths.
Under these conditions, the picture of synchronously rotating dipoles (with
constant phase difference relative to the field) breaks down. Instead, one
observes a mixture of rotating and counter-rotating or resting particles, as
our analysis of various angle distributions reveals. The desynchronization
induces, at the same time, a breakdown of the translational, layered structure.
Despite this complex many-particle behavior, we have shown that the boundary
can be well described in terms of the critical frequency $\omega_c(B_0)$ that
arises from a bifurcation in an effective {\em single-particle} approach for
the rotational motion in a viscous medium. This indicates that the appearance
of the high-frequency boundary is essentially a friction-induced effect.

A similar frequency-induced desynchronization effect has recently been
discussed by H\"artel {\em et al.},\cite{towing} who investigated a system of
interacting elongated particles in a rotating electric field via dynamic
density functional theory. Assuming a constant number density, the important
dynamic variable within the density functional approach is the orientational
distribution as function of time. At low and very high frequencies, the
distribution behaves similar to our distribution $f$ in that there is either a
single peak (reflecting synchronized motion with a constant phase difference)
or no peak at all. In the transition regime, however, H\"artel {\em et al.}
detected various new dynamic states characterized by time-dependent
oscillations and splitting of the peak in the distribution as well as an
overtaking by the driving field. In the present study we did not observe such
states, not even when looking at the time-dependence of our orientational
distributions (or the magnetization). It remains to be investigated whether
these qualitative differences in the rotational motion of anisotropic
many-particles systems are just due to differences in the specific model
system, or due to the fact that our results are based on a microscopic approach
rather than on the density field approach used in Ref.~\citenum{towing}.
Indeed, the relation between the microscopic and mesoscopic dynamics in driven
systems is an issue also discussed in other, related contexts, such as the
shear-induced dynamics of nanorods.\cite{Tao2009}

Finally, it is worth to briefly comment on the relevance of our dimensionless
model parameters in the context of real systems. The equilibrium parameters
considered here (density $\rho^{*}=0.1$, dipolar coupling strength $\lambda
\approx 6.7$) correspond to those of a strongly coupled ferrofluid exhibiting
chain formation.\cite{butter} Regarding the driving field, however, most of our
dimensionless frequencies $\omega_0^{*}$ are probably beyond the currently
accessible range. In many experiments involving rotating fields, the size of
the (typically superparamagnetic) particles considered is about $1$
$\mu$m.\cite{casic, elsner} A driving frequency of $\omega_0^* = 10$ (which is
well within the layered domain) then corresponds to an actual frequency of
about $10$ kHz if we assume room temperature ($T = 293$ K) and a mass density
of $5$ g/cm$^3$. This is $1$-$2$ orders of magnitudes larger than the
frequencies used in the literature.\cite{elsner,casic} Ferrocolloidal
particles, which have permanent dipoles (such as the ones considered here), are
often much smaller with sizes of about $10$ nm. In that case, $\omega_0^* = 10$
corresponds to a driving frequency of about $1$ GHz.

These considerations suggest that realistic driven systems will be fully
synchronized and layered according to our ``phase'' diagram in Fig.
\ref{fig:wEbd}. We note, however, that the actual location of the
desynchronization line encountered upon increasing $\omega_0^{*}$ depends on
the friction constant used in our BD simulations; \textit{i.e.}, increasing the
friction constant shifts the line towards lower frequencies (consistent with
the single-particle theory). Moreover, we have neglected in our study the
many-particle character of the hydrodynamic interactions induced by the
solvent. We would expect these interactions to effectively increase the
friction and thus shift the boundary towards even lower frequencies. Clearly,
it would be very interesting to actually incorporate such interactions by using
refined simulation methods such as, \textit{e.g.}, stochastic rotation
dynamics.\cite{Malevanets1999} Hydrodynamic interactions may also be relevant
to better explore phenomena such as chain-to-cluster transitions that have been
revealed by recent studies.\cite{casic} These issues, as well as the dynamic
behavior in even more complex field geometries, will be the subject of future
studies.

\section{Acknowledgements}
We gratefully acknowledge financial support from the DFG within the research
training group RTG 1558 {\em Nonequilibrium Collective Dynamics in Condensed
Matter and Biological Systems}, project B1.

\footnotesize{
\bibliography{ref}
\bibliographystyle{rsc}
}

\end{document}